\documentstyle[11pt,moriond,epsfig]{article}

\def\Journal#1#2#3#4{{#1} {\bf #2}, #3 (#4)}

\def\APJ{\em Astrophys. J.}
\def\APP{\em Astropart. Phys.}
\def\JPG{{\em J. Phys.} G}

\def\NIMA{{\em Nucl. Instr. Methods} A}

\def\PRD{{\em Phys. Rev.} D}

\def\la{\mathrel{\mathchoice {\vcenter{\offinterlineskip\halign{\hfil
$\displaystyle##$\hfil\cr<\cr\sim\cr}}}
{\vcenter{\offinterlineskip\halign{\hfil$\textstyle##$\hfil\cr<\cr\sim\cr}}}
{\vcenter{\offinterlineskip\halign{\hfil$\scriptstyle##$\hfil\cr<\cr\sim\cr}}}
{\vcenter{\offinterlineskip\halign{\hfil$\scriptscriptstyle##$\hfil\cr<\cr
\sim\cr}}}}}

\begin{document}

\pagestyle{plain}
\thispagestyle{empty}
\footskip 1cm
\renewcommand{\thefootnote}{\fnsymbol{footnote}}

\begin{center}

\vspace*{4cm}
{\Large\bf FIRST RESULTS FROM \\[.8ex]
THE KASCADE AIR-SHOWER EXPERIMENT \footnote{Talk given at {\sl The 
XXXIInd Rencontre de Moriond, ``Very High Energy Phenomena in 
the Universe''}, Les Arcs, France (1997)}}

\vspace*{1cm}

K.-H.~Kampert$^{a,b}$, 
W.D.~Apel$^{a}$, 
K.~Bekk$^{a}$, 
E.~Bollmann$^{a}$,
H.~Bozdog$^{c}$,
I.M.~Brancus$^{c}$,
M.~Brendle$^{d}$,
A.~Chilingarian$^{e}$,
K.~Daumiller$^{b}$, 
P.~Doll$^{a}$, 
J.~Engler$^{a}$, 
M.~F\"oller$^{a}$, 
H.J.~Gils$^{a}$,
R.~Glasstetter$^{a}$, 
A.~Haungs$^{a}$, 
D.~Heck$^{a}$, 
J.~H\"orandel$^{a}$, 
H.~Keim$^{a}$, 
J.~Kempa$^{c}$,
H.O.~Klages$^{a}$, 
J.~Knapp$^{b}$, 
H.J.~Mathes$^{a}$, 
H.J.~Mayer$^{a}$, 
H.H.~Mielke$^{a}$, 
D.~M\"uhlenberg$^{a}$, 
J.~Oehlschl\"ager$^{a}$, 
M.~Petcu$^{c}$, 
U.~Raidt$^{d}$, 
H.~Rebel$^{a}$, 
M.~Roth$^{a}$, 
G.~Schatz$^{a}$, 
H.~Schieler$^{a}$, 
G.~Schmalz$^{a}$, 
T.~Thouw$^{a}$, 
J.~Unger$^{a}$,
B.~Vulpescu$^{c}$, 
G.J.~Wagner$^{d}$, 
J.~Weber$^{a}$, 
J.~Wentz$^{a}$, 
T.~Wibig$^{f}$, 
T.~Wiegert$^{a}$, 
D.~Wochele$^{a}$, 
J.~Wochele$^{a}$, 
J.~Zabierowski$^{f}$, 
S.~Zagromski$^{a}$, and
B.~Zeitnitz$^{a,b}$\\[1ex]

--- KASCADE Collaboration ---

\vspace*{5mm}
\noindent
{\small\em
$^{a}$ Institut f\"ur Kernphysik, Forschungszentrum Karlsruhe, 
           D-76021 Karlsruhe, Germany, \\
$^{b}$ Institut f\"ur Experimentelle Kernphysik,
           Universit\"at Karlsruhe, D-76021 Karlsruhe, Germany,\\
$^{c}$ Institute of Physics and Nuclear Engineering,
           RO-7690 Bucharest, Romania, \\
$^{d}$ Physikalisches Institut, Universit\"at T\"ubingen,
           D-72076 T\"ubingen, Germany, \\
$^{e}$ Cosmic Ray Division, Yerevan Physics Institute, 
           Yerevan 36, Armenia, \\
$^{f}$ Inst.\ for Nuclear Studies and Dept.\ of
          Exp. Physics, University of Lodz,
          PL-90950 Lodz, Poland
}

\end{center}

\vspace*{1cm}

\abstracts{
A new extensive air shower (EAS) experiment has been 
installed at the laboratory site of the Forschungszentrum 
Karlsruhe.  The major goal of the experiment is to determine the 
chemical composition in the energy range around and above the 
knee of the primary cosmic ray spectrum.  An important advantage 
of the installation is the capability to simultaneously 
measure the electromagnetic, muonic and hadronic components of 
EAS event-by-event, thereby reducing systematic uncertainties to 
a large extend.  Data taking with a large part of the experiment 
has started at the end of 1995 with further installations 
continuing during 1996.  First preliminary results are 
presented.}

\section{INTRODUCTION}

Despite the fact that ultra-high energy cosmic rays (UHE-CR) are 
known for decades, their sources and the acceleration mechanism 
are still under debate.  Mainly for reasons of the required power 
the dominant acceleration sites are generally believed to be 
supernova remnants in the Sedov phase.  Naturally, this leads to 
a power law spectrum as is observed experimentally.  Detailed 
examination suggests that this process is limited to $E/Z \la 
10^{15}$\,eV. Curiously, the CR spectrum steepens at approx.\ $5 
\times 10^{15}$\,eV, indicating that the `knee' may be related to 
the upper limit of acceleration.  A change in the CR propagation 
with decreasing galactic containment has also been considered.  A 
key observable for understanding the origin of the knee and 
distinguishing the SN acceleration model from other proposed 
mechanisms \cite{szabo-94}, is given by the mass composition of 
CR particles and by possible variations across the knee.  
Unfortunately, beyond the knee little is known about the CR's 
other than their energy spectrum.  Direct measurements using 
detector systems on satellites, space craft or high altitude 
balloons cannot provide the required data with sufficient 
statistics because of limited detector area and exposure time 
\cite{mueller-91,ichimura-93}.  Experiments at ground level do 
not suffer from these problems since they can cover large areas, 
measure for extended time periods, and also take advantage of the 
magnifying effect of the atmosphere.  The latter is because 
sufficiently high energy CR particles initiate extensive air 
showers which spread out over large areas at observation level.  
Sampling detector systems with typical coverages of less than one 
percent can be used for registration of such showers.  However, 
this indirect method of detection bears a number of serious 
difficulties in the interpretation of the data and requires 
detailed modeling of the shower development and detector 
responses.  It is well known, that a number of characteristics of 
EAS depends on the energy per nucleon of the primary nucleus, 
notably the ratio of electron to muon numbers, the energy of the 
hadrons in the shower, the shapes of the lateral distributions of 
the various components of the shower, etc.\ The basic concept of 
the KASCADE experiment \cite{klages-96} is to measure a large 
number of these parameters for each individual event in order to 
determine both the energy and mass of the primary particles.

\section{LAYOUT AND STATUS OF THE EXPERIMENT}

\begin{figure}[tb]
\begin{minipage}[t]{7.7cm}
\begin{center}
\epsfig{file=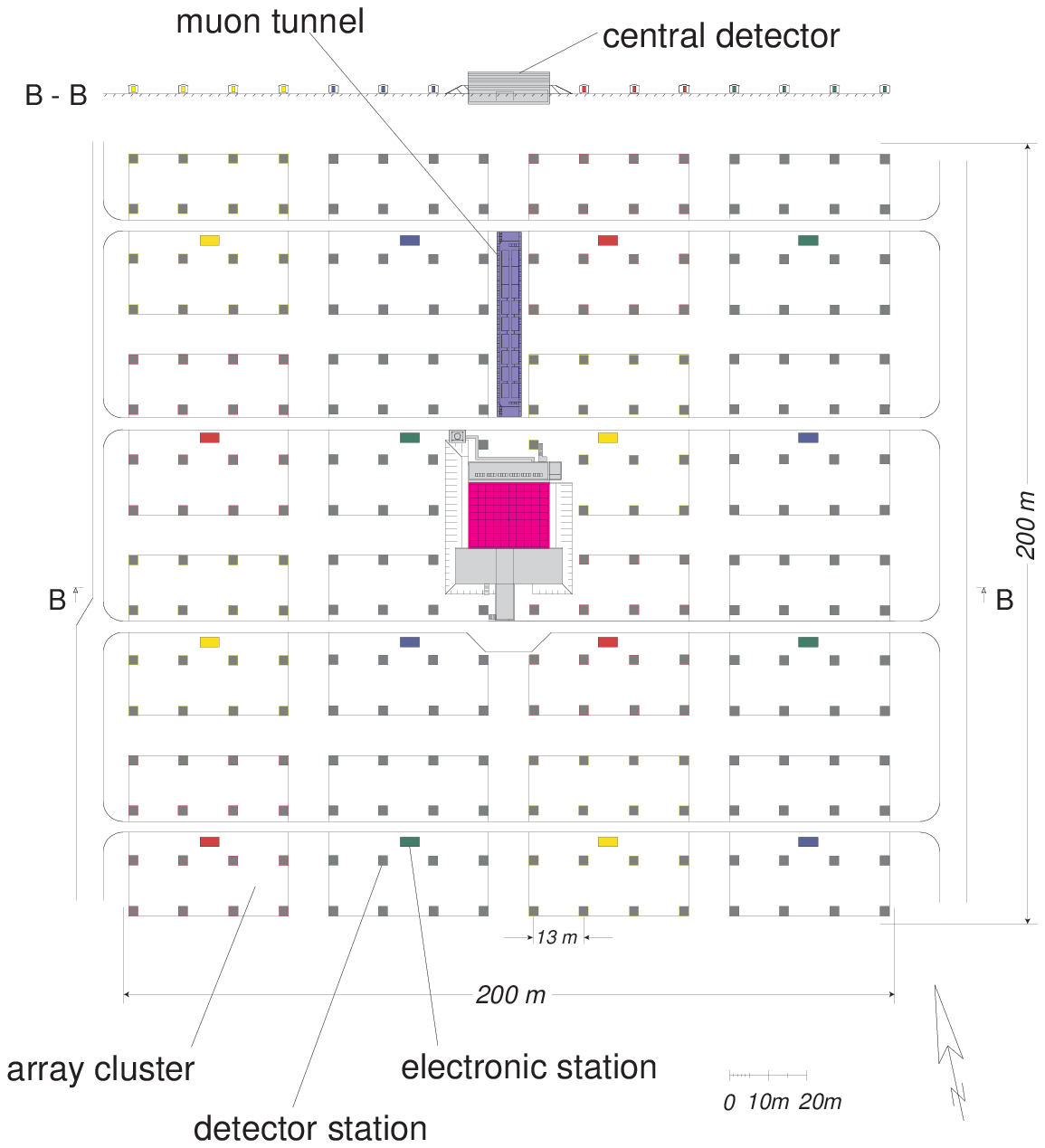,width=7.7cm}
\caption{Schematic layout of the KASCADE experiment.}
\label{fig:kascade-layout}
\end{center}
\end{minipage}
\hfill
\begin{minipage}[t]{7.7cm}
\begin{center}
\epsfig{file=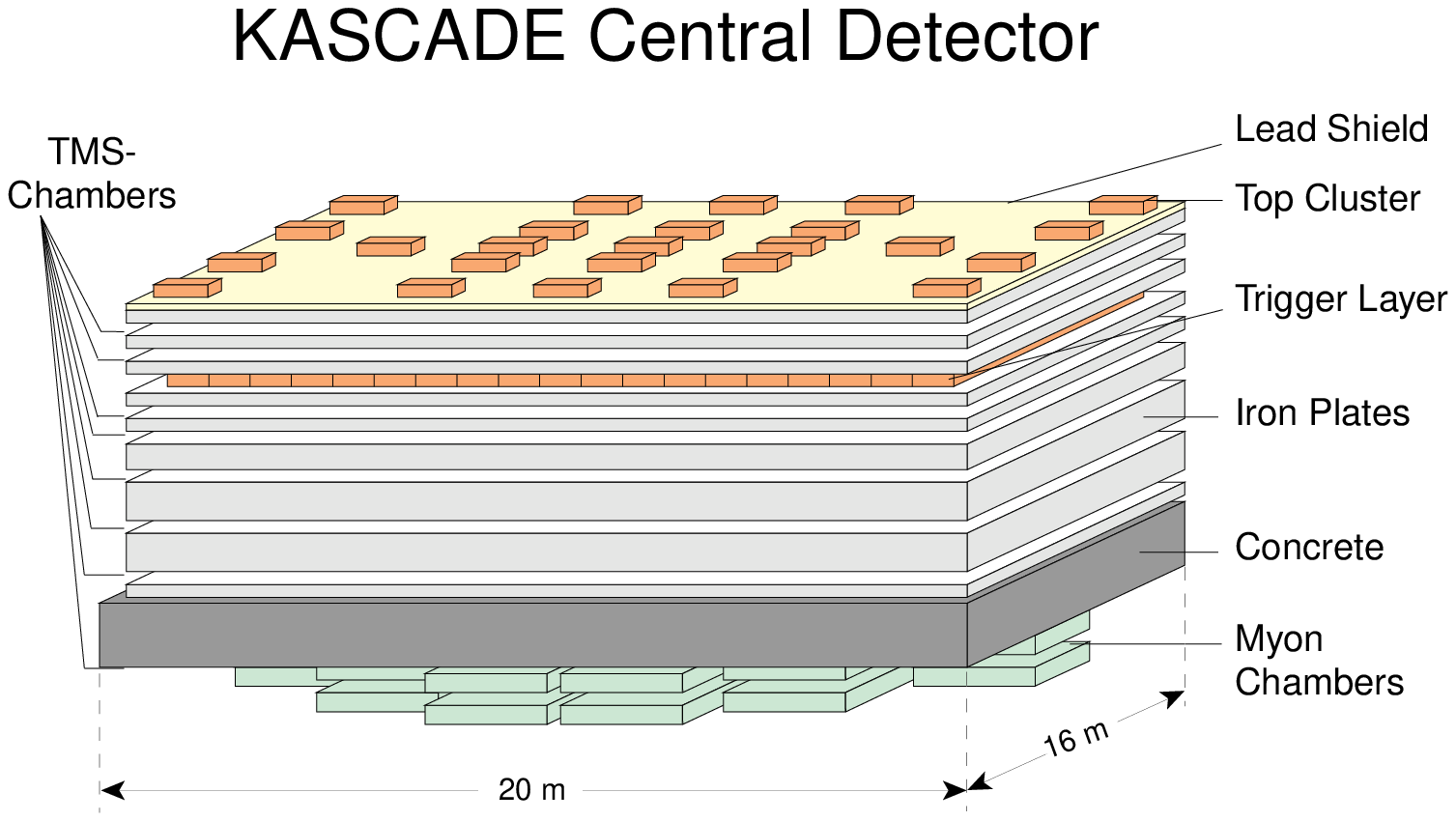,width=7.7cm}
\caption{Schematic layout of the central detector system of KASCADE.}
\label{fig:central-det}
\end{center}
\end{minipage}
\end{figure}

KASCADE (\underline{Ka}rlsruhe \underline{S}hower 
\underline{C}ore and \underline{A}rray \underline{De}tector) is 
located on the laboratory site of the Forschungszentrum 
Karlsruhe, Germany (at $8^{\circ}$ E, $49^{\circ}$ N, 110 m 
a.s.l.).  It consists of three major components; a scintillator 
array, a central detector system with a hadron calorimeter, and a 
large area muon tracking \cite{klages-96}.  Its schematic layout 
is shown in Fig.\,\ref{fig:kascade-layout}.

Scintillation detectors for the measurement of electrons and 
photons, and of muons outside the core region of extensive air 
showers are housed in 252 detector stations on a rectangular grid 
of 13 m spacing forming a detector array of $200 \times 200$ 
m$^{2}$.  Each station contains 2 or 4 scintillation detectors 
for the electron/photon component with a total area of (1.6 
m$^{2}$) 3.2 m$^{2}$.  Energy deposits equivalent to 2000 m.i.p.\ 
can be detected with a threshold of 0.25 m.i.p.  The 3.2 m$^{2}$ 
muon detector of a station is located below a shield of 20 
$X_{0}$ thickness.  It consists of 4 pieces of plastic 
scintillator, $90 \times 90 \times 3$ cm$^{3}$ each, read out by 
green wavelength shifter bars on all edges with a total of 4 
phototubes.  The supply and electronic readout of the detectors 
in the stations is organized in 16 clusters of 16 stations each 
(see Fig.\,1).  These clusters act as small independent air 
shower arrays.

The main part of the central detector system 
(Fig.\,\ref{fig:central-det}) is the finely segmented hadron 
calorimeter.  It consists of a $20 \times 16$ m$^{2}$ iron stack 
arranged into 9 horizontal slabs with a total absorber thickness 
corresponding to more than 11 nuclear interaction lengths.  A 
total of 10,000 ionization chambers filled with the room 
temperature liquid tetramethylsilane (TMS) is used for the 
measurement of energy in the gaps \cite{mielke-calo}.  The 
chambers are made of $50 \times 50 \times 1$ cm$^{3}$ stainless 
steel boxes and are read out by 4 independent central electrodes 
of $25 \times 25$ cm$^{2}$ size.  Thus, the readout of the 
calorimeter amounts to 40,000 electronic channels spread out over 
2,500\,m$^{2}$.  The fine segmentation allows for a separation of 
hadrons with distances as low as 50 cm.  The detectors have a 
dynamic range of $10^{4}$ limited only by the amplifier chain and 
its performance ensures a very stable operation over many years.  
The top layer of ionization chambers is shielded against the 
electromagnetic component of EAS by 12 cm of iron and additional 
5 cm of lead.  The energy sum of the hadrons in the core of a 1 
PeV shower can be determined with a resolution of 8\,\%.
Individual hadrons with energies larger than 20 GeV are 
reconstructed.  A prototype of the calorimeter has shown stable 
operation for about 3 years and has provided interesting results 
\cite{mielke-jpg,korn-jpg}.  In the third gap from the top of the 
iron stack (shielded by $\sim 30 X_{0}$) a layer of 456 
scintillation detectors is placed to trigger the readout of the 
calorimeter and other components of the experiment and to measure 
the arrival times of hadrons.  In addition, the trigger layer 
acts as a muon detector, allowing to determine the lateral- and 
time distributions of muons above a threshold of about 0.4 GeV. 
Underneath the calorimeter, two layers of multiwire proportional 
chambers (MWPCs) are used to measure muon tracks above an energy 
threshold of 2 GeV with an angular accuracy of about 
$1.0^{\circ}$.

North of the central detector a 50 m long and 5.5 m wide tunnel 
has been added to the experiment.  In this tunnel $600$ m$^{2}$ 
of limited-streamer tubes will be used in three layers for 
tracking of muons under a shielding of 18 $X_{0}$, corresponding 
to an energy threshold of 0.8 GeV. The tracking accuracy will be 
around $0.5^{\circ}$.  The detector \cite{doll-95} will have an 
effective area of about 150\,m$^{2}$ for the determination of the 
size and lateral distribution of the muon component and will -- 
for sufficiently large showers -- enable approximate 
determination of the mean muon production height by means of 
triangulation.  Shower simulations with the CORSIKA 
\cite{corsika} code show, that this observable provides another 
piece of information about the mass of the primary particle.

Data taking has started in late 1995 with large parts of the 
experiment in stand-alone mode.  Correlated data are being 
collected since April 1996 with 100 m$^{2}$ area of the 
calorimeter operational at the beginning.  By now the read-out 
area has been doubled.  The remaining part of the calorimeter and 
the muon tunnel are expected to start operation in 1997.  At 
present, trigger thresholds of the array, trigger plane, and 
top-cluster are adjusted to limit the total trigger rate to $\sim 
1$ Hz.  In the near future this number will be increased to 
approx.\ 10 Hz thereby lowering the effective energy threshold to 
$\sim 10^{13}$ eV for specific events.

\section{PRELIMINARY RESULTS}

Owing to the ongoing installation and calibration work during 
normal working hours and data taking during nights and over 
weekends only, analysis is still in a preliminary stage and only 
a small fraction of the data available could be analyzed.  
Typically, about half a million of showers per week are collected 
during this mode of operation.

\begin{figure}[tb]
\begin{minipage}[t]{7.7cm}
\begin{center}
\epsfig{file=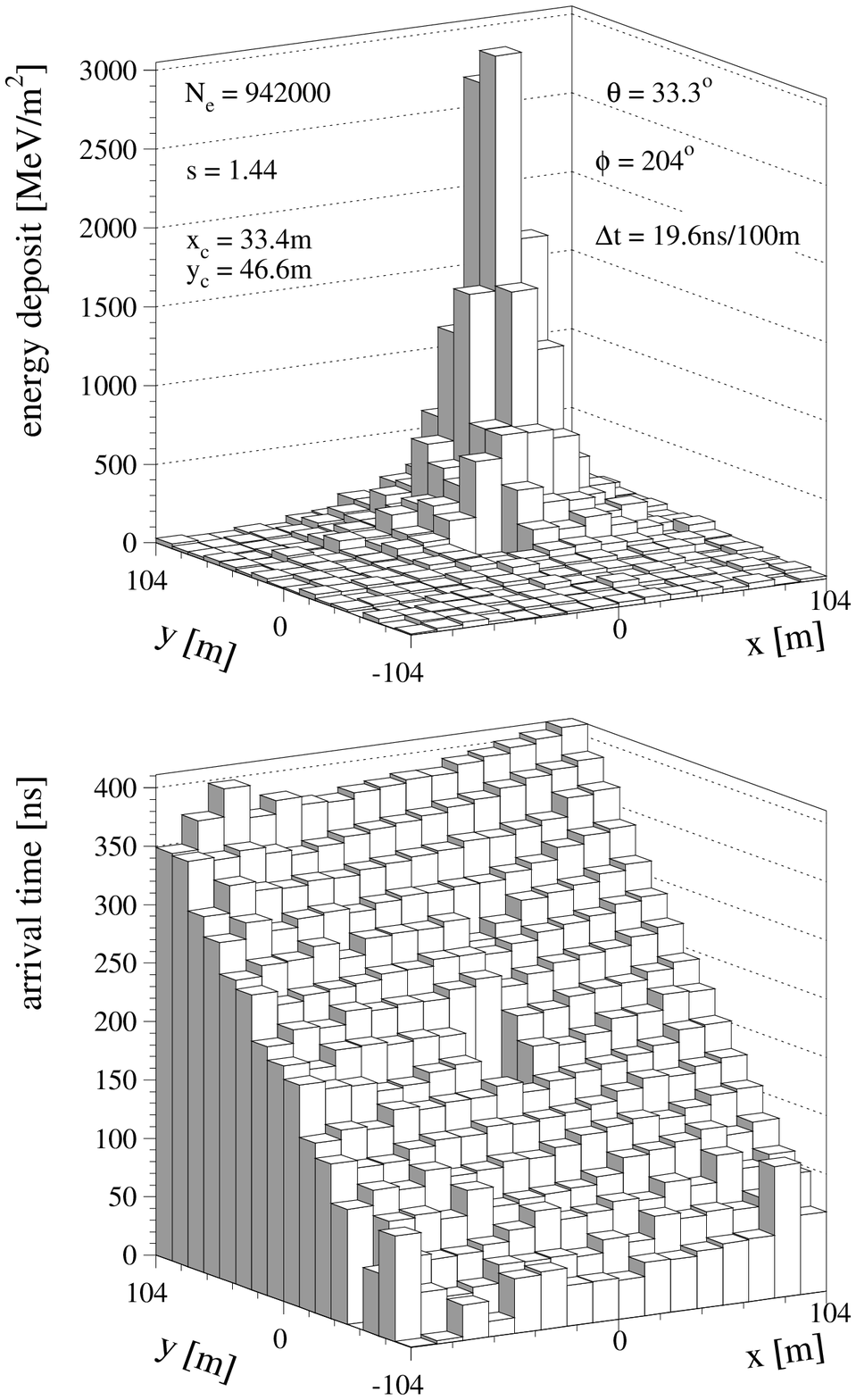,width=7.7cm}
\caption{Example of an EAS as observed by the electron/gamma 
detectors of the array. In the two parts, each column represents
the energy deposit and the arrival time of the shower front, 
respectively. Reconstructed shower parameters are listed.}
\label{fig:array-event}
\end{center}
\end{minipage}
\hfill
\begin{minipage}[t]{7.7cm}
\begin{center}
\epsfig{file=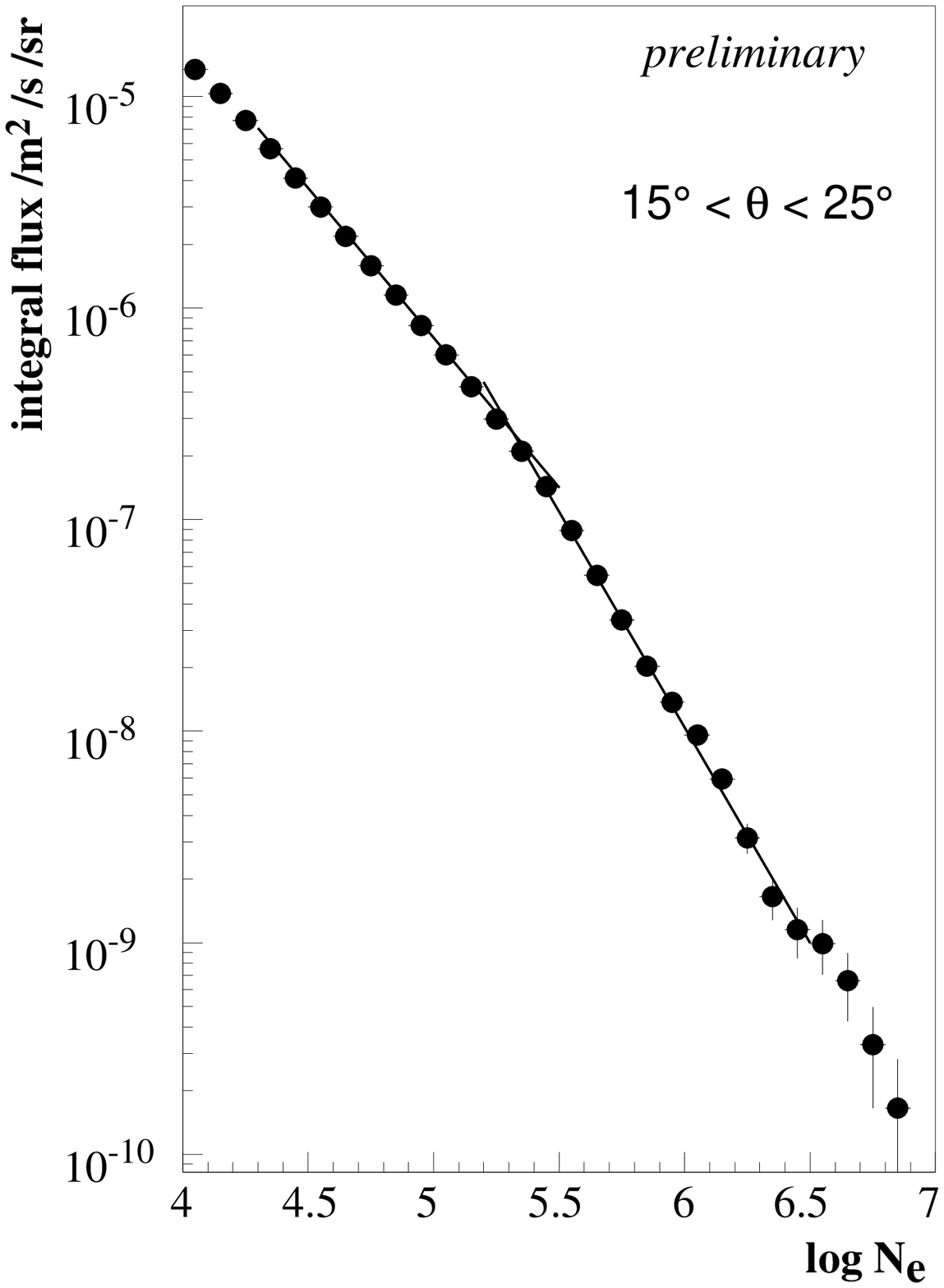,width=7.7cm}
\caption{Event rate as a function of reconstructed shower size 
for EAS in the zenith angle range $15^{\circ}$ to $25^{\circ}$.}
\label{fig:knee}
\end{center}
\end{minipage}
\end{figure}

Figure \ref{fig:array-event} shows as an example a shower as 
registered by the electron/gamma detectors of the array.  In the 
upper part the deposited energies and in the lower part the 
corresponding arrival times of the same event are shown for each 
detector station.  Parameters resulting from the analysis such as 
electron number $N_{e}$, core position, shower direction, and age 
are given in the figure.  It is obvious from the distributions 
that these numbers can be determined to high accuracy.  The 
lateral distribution of all charged particles in the shower is 
analyzed using a NKG function which fits the data nicely within 
the typical range of our measurement, say 10-200 m.  The muon 
signals are analyzed in a similar way, however, due to 
electromagnetic punch through close to the shower core, this 
measurement is at present restricted to radii larger than 40 m.  
When we plot the event rate as a function of the reconstructed 
electron size of the showers in the zenith angle range 
$15^{\circ}$ to $25^{\circ}$ (see Fig.\,\ref{fig:knee}) we find a 
distinct change in the slope near $\log N_{e} = 5.5$.  These data 
correspond to only one week of measurement.  Both, the position 
of the break in the spectrum as well as the spectral indices 
below and above the knee are well in agreement with expectations 
from earlier experiments.  Presently, these data are analyzed in 
much more detail to carefully correct for effects like trigger 
thresholds, array efficiency, trigger biases, etc.\ and to 
investigate their systematic uncertainties.

Simultaneous reconstruction of the electron- and muon size will 
enable us to determine their ratio event-by-event.  This provides 
a parameter which has proven large sensitivity to the primary 
composition.  Details of the electron and muon lateral 
distributions are also studied to get a better understanding of 
the basic properties of air showers.  The influence of 
atmospheric parameters to the ground level observables of EAS has 
been calculated and the results will be checked in the ongoing 
analysis.

\begin{figure}[tb]
\begin{minipage}[t]{7.7cm}
\begin{center}
\epsfig{file=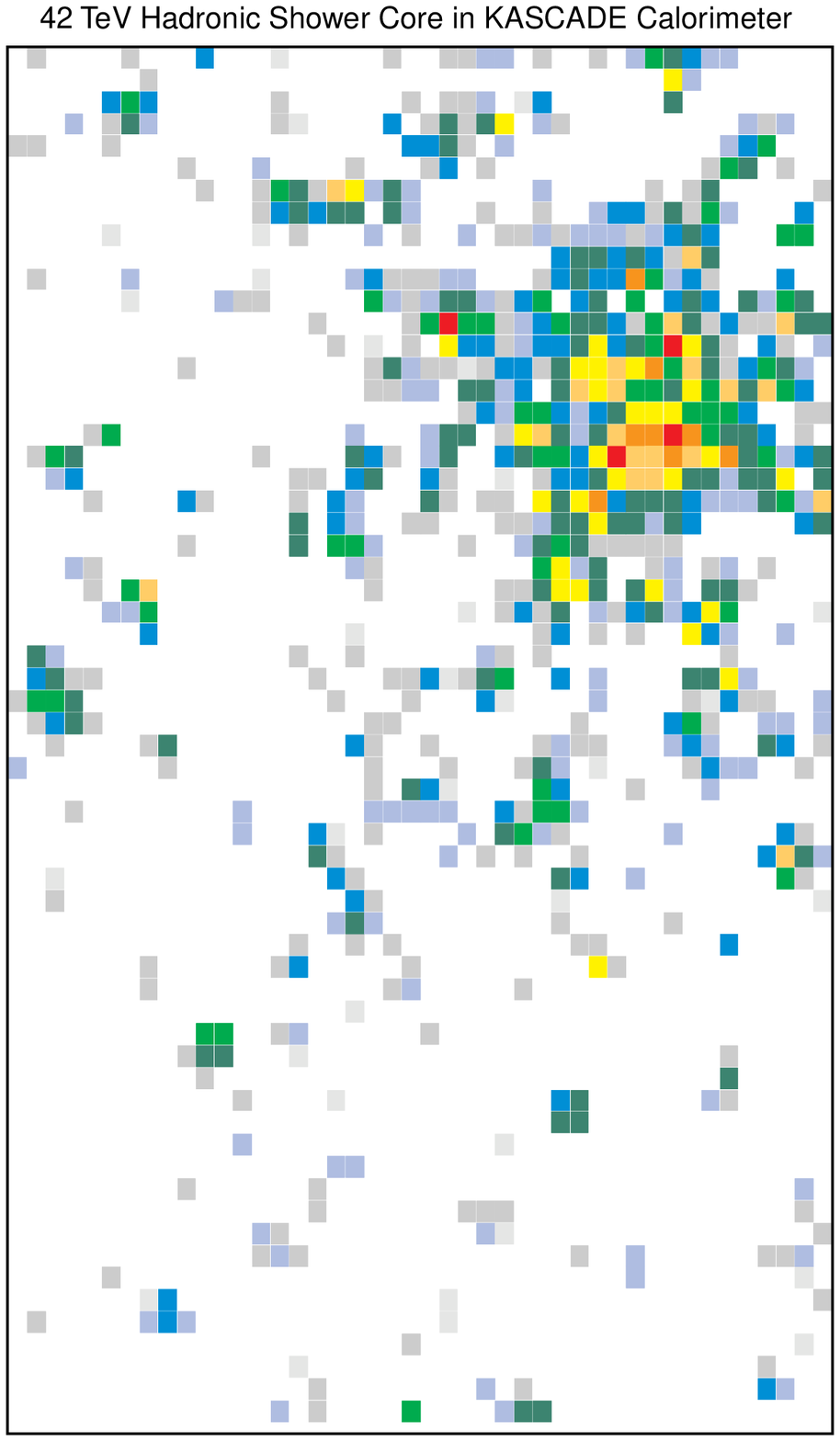,width=6.5cm}
\caption{Pattern of signals for a shower core as seen in the
3$^{\rm rd}$ layer of the calorimeter. Each box represents the
signal of an individual ionization chamber, i.e.\ the total size
of the figure corresponds to $10 \times 16$ m$^{2}$. The energy 
of the shower corresponds to $3\cdot10^{15}$ eV.}
\label{fig:calo-event}
\end{center}
\end{minipage}
\hfill
\begin{minipage}[t]{7.7cm}
\begin{center}
\epsfig{file=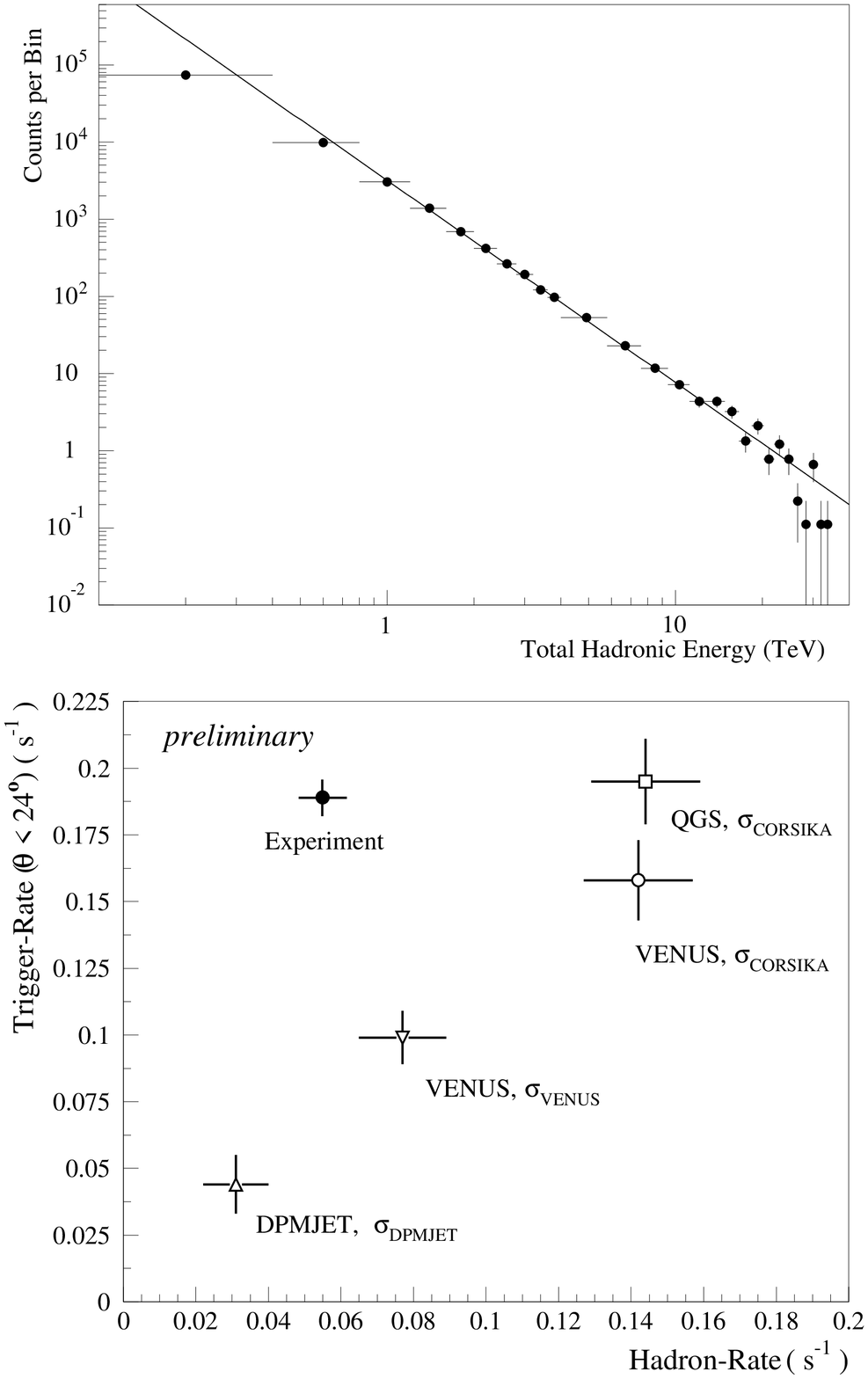,width=7.0cm}
\caption[xx]{Top:) Distribution of the total hadronic energy per
EAS above the trigger threshold \cite{unger-dr}. The line shows a
power-law fit with an index of $-2.62 \pm 0.09$.
Bottom:) Observed experimental trigger- and hadron rate in 
comparison to expectations from different hadronic interaction
models (see text for details).}
\label{fig:hadron-spec}
\end{center}
\end{minipage}
\end{figure}

A unique feature of KASCADE is its large hadronic calorimeter.  
Measuring the properties of the shower core along with the 
information from the electron and muon detectors at larger 
distances will substantially improve the sensitivity to the 
chemical composition particularly at energies below the knee.  
Furthermore, the low trigger threshold of approx.\ $10^{13}$\,eV 
for proton induced showers provides a window of significant 
overlap with direct CR measurements where both the energy- and 
mass spectrum are known to a reasonable precision.  The comparison 
of EAS and direct measurements will thus serve as a vital test of 
the hadronic interaction (and atmospheric shower propagation) 
models.  Even in the energy range accessible to operating 
colliders, up to an equivalent laboratory energy of 
$1.6\cdot10^{15}$\,eV, our knowledge of hadronic interactions is 
very limited because collider experiments do not cover the very 
forward fragmentation region.  This region is most important to 
model the hadronic development of EAS and it can be studied by 
using the calorimeter data itself.  Similarly, very energetic 
single hadrons not accompanied by an air shower can reach ground 
level after only very few hadronic interactions.  Again, 
their rate as a function of energy is very sensitive to the 
average hadronic inelasticity and to the inelastic cross section.  
Here, we report only on very first data on hadronic shower cores 
\cite{unger-dr}.

Figure \ref{fig:calo-event} shows as an example a two-dimensional 
pattern of energy deposition in the 3$^{\rm rd}$ layer of the 
calorimeter.  This information is available for each of the 8 
readout planes and is used to identify hadron tracks through the 
calorimeter.  Simulations show that the tracking efficiency for 
$E_{h} \ge 100$\,GeV is larger than 90\,\% of for shower energies 
below $10^{15}$ eV \cite{unger-dr}.  In the event shown, the 
total hadronic energy, as measured by the calorimeter, was 42 TeV 
and the energy of the primary particle, as reconstructed from the 
array data, was about $3\times10^{15}$\,eV. The multiplicity of 
reconstructed hadrons in some events exceeds 100 and the total 
hadronic energy sum spectrum is plotted in 
Fig.\,\ref{fig:hadron-spec} (top).  The data are well described 
by a power-law with an index close to that of the primary 
spectrum.  The lateral distribution of hadrons is found to 
flatten with increasing shower size.  In order to quantitatively 
compare the observed experimental trigger and reconstructed 
hadron rates with model predictions, detailed simulations have 
been performed by using composition and energy distributions of 
primary CR's as input.  These simulated data were fed through the 
same chain of programs as used for real data and some of the 
results are shown in Fig.\,\ref{fig:hadron-spec} (bottom).  The 
still preliminary analysis indicates that even small variations 
of the inelastic cross section used in the model calculations 
lead to strong effects in the expected rates.  Furthermore, none 
of the models seems to be able to reproduce both the trigger rate 
(multiplicity $\ge$ 6 in the trigger layer) and the reconstructed 
hadron rate simultaneously.  Although, more detailed studies are 
required to settle this problem, the data demonstrate the power 
to experimentally test hadronic interaction models in the very 
forward region.

\section{CONCLUSIONS}

The installation of the KASCADE experiment is nearing completion 
and data taking with a large part of the experiment has started.  
Preliminary data of the array and central detector look very 
promising.  No reliable results of composition can be given at 
the time of writing this report.  Mandatory for such an analysis 
is to prove that {\em (i)} the detector and reconstruction 
methods are understood in detail, and {\em (ii)} the hadronic 
interaction models and atmospheric shower simulations are well 
under control.  An example of such kind of analysis has been 
discussed briefly.  In parallel, multiparameter analyses are 
under development to infer the composition from the different 
components of the experiment.

\end{document}